\documentclass[12pt,preprint]{aastex}

\begin{document}
\title{The Misalignments between Matter and Galaxy Distributions in 
Triaxial Clusters: A Signature of a Possible Fifth Force?}
\author{Jounghun Lee}
\affil{Department of Physics and Astronomy, FPRD, Seoul National University, 
Seoul 151-747, Korea} 
\email{jounghun@astro.snu.ac.kr}
\begin{abstract}
The standard structure formation model based on a $\Lambda$CDM cosmology predicts 
that the galaxy clusters have triaxial shapes and that the cluster galaxies have a 
strong tendency to be located preferentially along the major axes of host cluster's dark 
matter distributions due to the gravitational tidal effect. The predicted correlations between 
dark matter and galaxy distributions in triaxial clusters are insensitive to the initial 
cosmological parameters and to the galaxy bias, and thus can provide a unique test-bed for 
the nonlinear structure formation of the $\Lambda$CDM cosmology.  
Recently, Oguri et al. determined robustly the dark matter distributions in the galaxy 
clusters using the two dimensional weak lensing shear fitting and showed that the 
orientations of the cluster galaxy distributions are only very weakly correlated with those of 
the underlying dark matter distributions determined robustly, which is in contrast to  with 
the $\Lambda$CDM-based prediction. 
We reanalyze and compare quantitatively the observational result with the 
$\Lambda$CDM-based prediction from the Millennium Run simulation with the help 
of the bootstrap resampling and generalized $\chi^{2}$-statistics. The hypothesis that 
the observational result is consistent with the $\Lambda$CDM-based prediction is 
ruled out at the $99\%$ confidence level. A local fifth force induced by a non-minimal 
coupling between dark energy and dark matter might be responsible for the observed 
misalignments between dark matter and galaxy distributions in triaxial clusters.  
\end{abstract}
\keywords{cosmology:theory --- large-scale structure of universe}

\section{INTRODUCTION}

Although the dark energy that is responsible for the present cosmic acceleration  
is known to occupy more than $70 \%$ of the energy budget of the Universe 
\citep[e.g.,][]{wmap7}, its nature is still shrouded in deep mystery. The 
simplest candidate for the dark energy is the cosmological constant ($\Lambda$) 
that Einstein introduced in his field equation a century ago and called {\it the 
fundamental constant of Nature} \citep{einstein17}, which has been interpreted 
as the vacuum energy that is completely homogeneous in space, does not evolve with 
time, non-interacting, chemically inert, and has a constant equation of state 
\citep[see][for a review]{carroll-etal92}. 

Over the past two decades, the cosmologists have witnessed and enjoyed the 
practical success of the $\Lambda$CDM (cosmological constant + cold dark matter) 
cosmology in explaining the large-scale properties of the Universe, which have 
left little doubt on the validity of the $\Lambda$CDM as a standard model. 
Nevertheless, an increasing number of literatures have been devoted to 
seeking for viable alternatives that could compete with and hopefully 
replace the $\Lambda$CDM \citep[][for a review]{Alcaniz06}. The main 
motivation of the alternative dark energy models is to explain naturally the 
cosmic coincidence that the densities of dark energy and dark matter have the 
same order of magnitude of unity, which issue the standard $\Lambda$CDM is 
incapable of addressing \citep{Amendola00,chiba01,chimento-etal03}.

An intriguing idea that was motivated by a fundamental particle physics and 
has gained a lot of popularity recently in the field of large scale structure is 
that dark energy is a scalar field having a non-minimal coupling with
dark matter \citep[][and references therein]{Amendola00,FP04,Alcaniz06}.  
The distinct feature of this fascinating idea is that a possible fifth force can 
be generated by the propagation of dark matter inhomogeneities into the 
scalar field due to the non-minimal coupling. The strength, range and potential 
form of a possible fifth force depends on how to model the coupling in the dark 
sector (dark energy+dark matter). 
The reason for the recent sharp attention on the coupled scalar field models 
is that the inclusion of a fifth force generated by the dark sector interaction  
may be able to explain several observational phenomena 
\citep{Maccio-etal04,nusser-etal05,PN10,baldi-etal10}, which have been 
suggested as mismatches to the theoretical predictions of the $\Lambda$CDM 
cosmology such as the abundance of galaxy satellites, the speed of bullet 
cluster, and the clear-out of void dwarfs 
\citep{klypin-etal99,peebles01,FR07,LK10}. 

In this Letter, we present another observable that exhibits an obvious mismatch  
with the numerical prediction based on the standard structure formation based on 
a $\Lambda$CDM cosmology: the misalignments between matter and galaxy 
distributions in triaxial galaxy clusters. In conventional approaches the shapes 
of dark matter distribution in galaxy clusters were often determined by measuring 
the member galaxy distributions \citep[e.g.,][]{binggeli82,flin87,rhee-katgert87,
west-etal95,chambers-etal02,PB02,hashimoto-etal08,EB09}. 
Recent progress in weak lensing shear analysis, however, has allowed us to 
determine the shapes of clusters directly from dark matter distributions 
\citep{okabe-etal10,Oguri-etal10}, which has revealed that the cluster dark matter 
distributions  are only very weakly aligned with the cluster galaxy distributions. 

Noting that the standard structure formation model based on $\Lambda$CDM cosmology 
predicts very strong correlations between the shapes of galaxy and dark matter 
distributions \citep{Lee-etal08}, we make a quantitative comparison between the 
observed trend and the $\Lambda$CDM-based prediction, and make a robust test of the 
hypothesis that the observed trend is consistent with the $\Lambda$CDM-based 
prediction with the help of the bootstrap resampling and generalized $\chi^{2}$ statistics.  We speculate that the observed misalignments between dark matter and galaxy 
distributions in triaxial clusters might be a new observational signature of a dark sector 
interaction. 

\section{NUMERICAL RESULT BASED A $\Lambda$CDM COSMOLOGY}

The correlations between the major axes of dark matter and satellite galaxy 
distributions in dark halos have been investigated by \citet{Lee-etal08} 
using the data from the mili-Millennium simulation that was run on a periodic 
box size of $62.5\,h^{-1}$Mpc for a flat $\Lambda$CDM cosmology with 
$\Omega_{m}=0.25,\ \Omega_{\Lambda}=0.75,\ h=0.73,\ \sigma_{8}=0.9,\ n_{s}=1$ 
\citep{springel-etal05}.
The dark halos and their substructures were identified from the mili-Millennium 
data by applying the Friends-of-Friends (FoF) and the SUBFIND algorithm, 
respectively \citep{springel-etal01}. The luminosities of the satellite galaxies 
residing in the dark halo substructures were determined according to 
the semi-analytic model of galaxy formation \citep{croton-etal06}. 
For a detailed description of the mili-Millennium simulation and the semi-
analytic galaxy catalog, we refer the readers to \citet{springel-etal05}.

Selecting those mili-Millennium halos which have five or more satellite galaxies, 
\citet{Lee-etal08} ended up having a sample of $277$ dark halos with mass 
$M\ge 1.3\times 10^{13}\,h^{-1}M_{\odot}$ at $z=0$. They determined the major 
axes of the selected $277$ halos using two different schemes. The first scheme is 
based on the dark matter distribution of each halo. They calculated the inertia 
momentum tensor of each halo using the constituent dark matter particles 
(each of which has mass $8.6\times 10^{8}\,h^{-1}M_{\odot}$) and determine the 
major axis as the eigenvector of the inertia momentum tensor corresponding to the 
largest eigenvalue. In the second scheme the major axis of the inertia 
momentum tensor of each halo was calculated by using the satellite galaxies 
weighted by their luminosity. 

The correlations between the major axes determined by the two different 
schemes were quantified in terms of the distribution of the alignment angle: 
Basically, they measured the angle between the two major axes in range of 
$[0,90^{o}]$ for each halo.  Binning the alignment angles and counting the 
number of halos whose alignment angles belong to each bin, they calculate 
the number fraction distribution as a function of the angle between the two 
major axes of dark matter and satellite galaxy distributions 
(see Figure 1 in Lee et al. 2008). Very strong correlations between the 
two major axes were found, which proved that the satellite galaxies 
in the massive halos tend to be located near the major axes of the dark 
matter distributions. 

The correlations between matter and galaxy distributions in halos are usually 
ascribed to the effect of the external tidal fields: According to the cosmic web theory 
based on a cold dark matter paradigm \citep{bond-etal96}, the coherence and non-linear 
sharpening of the gravitational tidal fields induce the  shape-alignments between the large-
scale structures such as cluster-supercluster alignments, void-supercluster alignments and 
the inclination of the galaxy spin axes onto the sheets 
\citep[e.g.,][]{navarro-etal04,KE05,atlay-etal06,trujillo-etal06,LE07,PL07}.

The gravitational tidal effects also cause the galaxy clusters to 
have triaxial shapes and  induce the correlations between matter and satellite 
galaxy distribution in the triaxial clusters since the merging and accretion of 
matter and galaxies onto the clusters occur preferentially along the primary 
filaments that are aligned with the directions of the minimal matter 
compression, i.e., the minor principal axes of the gravitational tidal fields 
\citep{west-etal95,bailin-etal05,BS05}.  After anisotropic merging along the primary 
filaments, however, the secondary infall and subsequent nonlinear evolution inside the 
clusters would modify the spatial distributions of the member galaxies \cite{dubinski99}. 
For instance, \citet{LK06} have shown that as the clusters become more relaxed, 
the correlations between the principal axes of dark matter and substructures 
tend to decrease. They also found that the correlations increase with redshift 
and cluster mass.

In observations, the correlations between the major axes of dark matter and 
galaxy distributions are often measured in the two dimensional projected planes. 
Therefore, for a direct comparison with the observational result, it is necessary to 
predict numerically the correlations of the projected major axes of dark matter and 
satellite galaxy distributions. Adopting the same mili-Millennium Run data used 
by \citet{Lee-etal08} and assuming the flat-sky approximation, we investigate 
here the correlations between the projected major axes of dark matter and 
galaxy distributions. Measuring and binning the angles $\phi$ of the projected 
major axes of dark matter and satellite galaxy distributions of the selected 
$277$ halos in the two dimensional plane, we find the number fraction of 
dark halos as a function of $\phi$.  
 
\begin{figure}
  \begin{center}
   \plotone{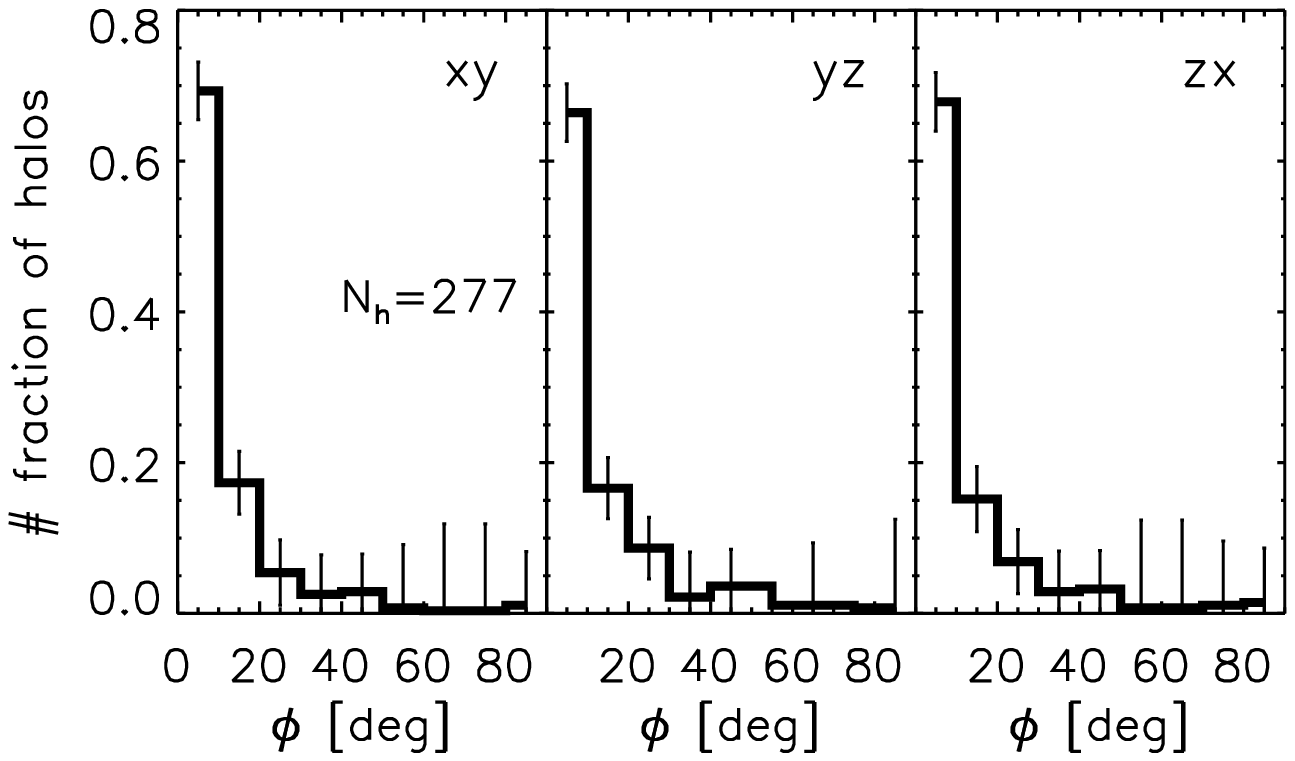}
\caption{The fraction of the dark halos as a function of the angles between the 
major axes of the cluster halos and the spatial distributions of the member 
galaxies projected into the $x$-$y$, $y$-$z$ plane and $z$-$x$ plane (in the 
left, middle and right panel, respectively).}
\label{fig:theory}
 \end{center}
\end{figure}

The left, middle and right panel of Figure \ref{fig:theory} plots the results 
for the case that the projection was done onto the xy, yz, and zx plane, 
respectively. The errors include both of the sample variance as well as the 
Poisson noise. The sample variance is calculated as one standard deviation 
among  the $1000$ bootstrap resamples. As one can see, the distributions has a 
sharp peak at the first bin $0^{o}\le\phi\le 10^{o}$, as in the three 
dimensional case, which proves that projections onto the two dimensional 
planes do not diminish the correlations between the major axes of matter 
and satellite galaxy distributions. In the following section, we compare 
this numerical prediction with the observational result and test the hypothesis 
that the observational result is consistent with the numerical prediction based 
on a $\Lambda$CDM cosmology.

\subsection{NUMERICAL VS. OBSERVATIONAL}

Performing two dimensional analysis of the weak lensing shear maps 
constructed with high-quality Subaru/Supreme-cam \citep{miyazaki-etal02},
\citet{Oguri-etal10} have measured the two dimensional projected shapes and 
orientations of dark matter distributions in the galaxy clusters from
the Local Cluster Substructure Survey \citep{okabe-etal10}. 
Using a subsample of $18$ galaxy clusters for which the weak lensing 
shear maps are well fitted to the elliptical density profiles \citep{JS02}, 
they determined the  longest axis of the dark matter distribution and 
measured its position angle for each cluster (see Table 1 in Oguri et al. 2010).

Selecting those member galaxies whose r-band magnitudes are brighter than $22$ 
mag, they also fitted the member galaxy distribution (confined to the same region) 
in each cluster to an elliptical power law profile and determined the positions angle 
of the longest axis of galaxy distributions (see Table 2 in Oguri et al. 2010). 
According to \citet{Oguri-etal10}, the fitting of the dark matter and 
member galaxy distribution in each cluster was done at the square region 
$20^{\prime}\times 20^{\prime}$, around cluster's center chosen 
to be the location of the brightest cluster galaxy, where $20^{\prime}$ 
corresponds to approximately $2$-$3$ Mpc for the selected clusters in a 
redshift range of $0.1< z < 0.3$. For a detailed description of the two dimensional 
fitting of weak lensing shear field and the cluster sample, see \citet{Oguri-etal10}.

By comparing the position angles of dark matter and galaxy distributions in each 
cluster, they found only very weak correlations between the two. They considered 
several possible systematics such as galaxy bias, dilution effect, fitting area and 
field-galaxy contamination but found no systematics significant enough to explain the 
observed trend. Although \citet{Oguri-etal10} asserted that the ellipticity distribution of 
dark matter distributions in the observed galaxy clusters agrees well with 
the theoretical prediction based on a $\Lambda$CDM cosmology \citep{JS02}, 
they did not recognize that the observed misalignments between the longest axes 
of dark matter and member galaxy distributions in the clusters are inconsistent 
the $\Lambda$CDM prediction. 

To make a parallel comparison of the observational result with the $\Lambda$CDM 
prediction given in \S 2,  here we determine the distributions of the angles $\phi$ 
between the projected major axes of dark matter and galaxy distributions using the 
observational data of the $18$ clusters given in \citet{Oguri-etal10}. The angle $\phi$ 
for each cluster is determined by  subtracting the position angle of the galaxy distribution 
from that of dark matter distribution and taking its absolute value. If the 
position angle difference exceeds $\pi/2$, then we take $\pi-\phi$. Then, 
we bin the values of $\phi$ and count the number of the clusters belonging to 
each bin.

\begin{figure}
  \begin{center}
   \plotone{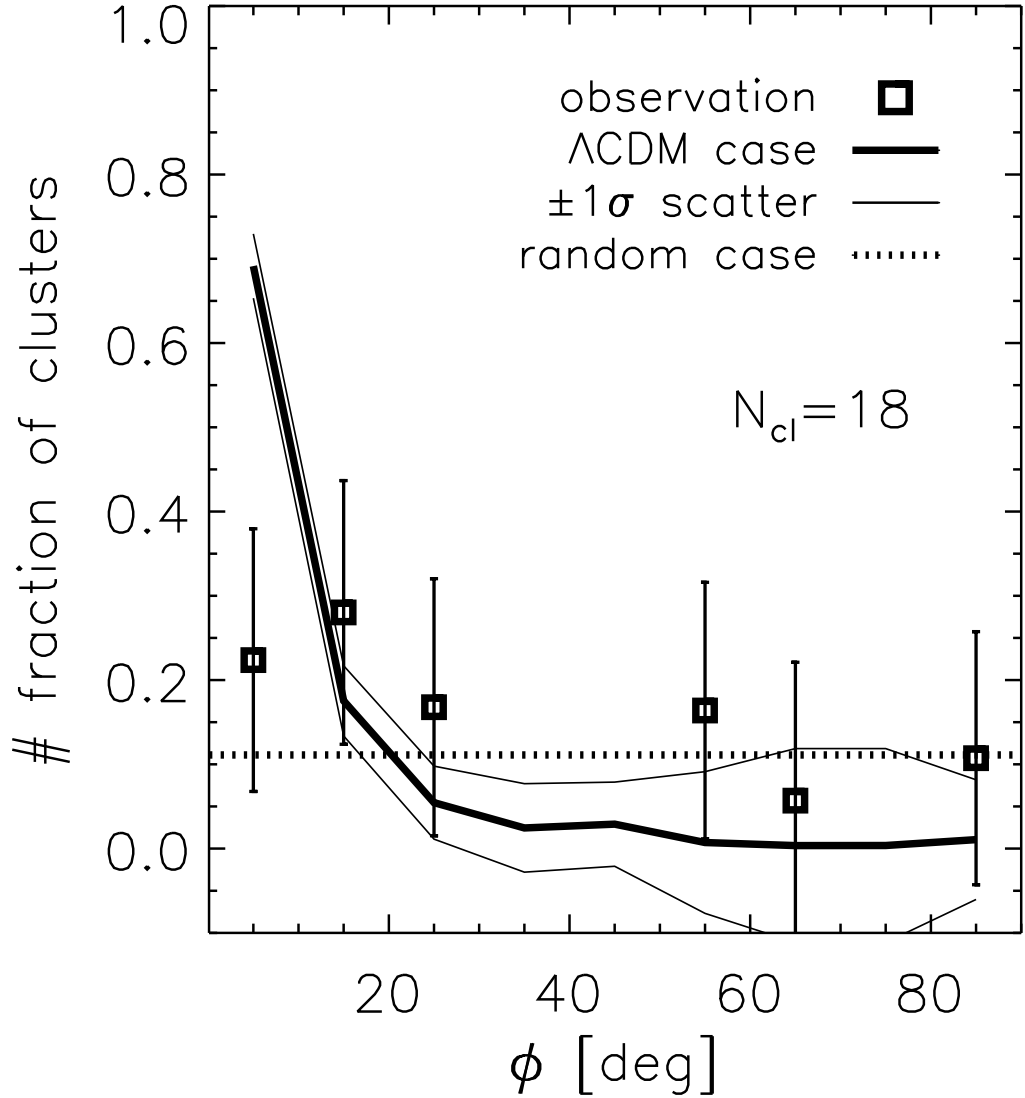}
\caption{Comparison of the $\Lambda$ prediction (solid line) with the 
observational result (open squares)obtained from Oguri et al. (2010).  The 
error-bars include the sample variance as well as the Poisson noise. The thin 
solid lines represent the $\pm 1\sigma$ bootstrap scatters.}
\label{fig:compare1}
 \end{center}
\end{figure}

Figure \ref{fig:compare1} plots the result as open squares. The dotted line 
represents the uniform distribution corresponding to the case that there is no correlation 
between the major axes of dark matter and galaxy distributions. The errors include the 
sample variance and the Poisson noise. As in \S 2, the sample variance is calculated as 
$1\sigma$ scatter among the $1000$ bootstrap resamples. For comparison, we also plot 
the numerical prediction of the $\Lambda$CDM cosmology obtained in \S 2 as thick solid 
line (projection onto xy plane) while the thin solid lines represent the $\pm 1\sigma$ 
bootstrap errors of the numerical results.  
  
As it can be seen, the observational result deviates significantly from the numerical 
result based on a $\Lambda$CDM prediction and the semi-analytic galaxy formation 
model. To quantify the deviation, we employ the generalized $\chi^{2}$-statistics which 
accounts for the correlations between the neighbor $\phi$-bins due to the errors in the 
measurement of cluster's position angles:
\begin{equation}
\label{eqn:chi2}
\chi^{2} = \Sigma_{i,j}\Delta f_{j}C^{-1}_{ij}\Delta f_{i} 
\end{equation}
Here $C_{ij}$ is the covariance matrix defined as 
$C_{ij}\equiv \langle\Delta f_{i}\Delta f_{j}\rangle$ and 
$\Delta f_{i}$ is the difference between the observed and predicted value of 
the cluster number fraction at the $i$-th bin given as 
$\Delta f_{i}\equiv f^{obs}_{i} - f^{the}(\phi_{i})$. 
The bracket in the definition of $C_{ij}$ (Equation \ref{eqn:chi2} represents 
the ensemble average over the $1000$ bootstrap resamples. 
Noting that in the large $\phi$ section (beyond the third bins) the Poisson 
noises dominate over the bootstrap errors and the signals of the numerical 
results are confined to the first two bins, 
we consider the first three bins ($0^{o}\le\phi\le 30^{o}$) for the calculation 
of $\chi^{2}$ (i.e., the degree of freedom is $3$).  

Putting the numerical results obtained in \S 2 into $f^{the}(\phi_{i})$ for the calculation 
of $\chi^{2}$, we test the hypothesis that the observational result is consistent with the 
$\Lambda$CDM prediction and found that the hypothesis is rejected at the $99\%$ confidence level. Meanwhile, putting the uniform distribution into $f^{the}(\phi_{i})$, we 
also test the null hypothesis that there is no correlation between the major axes of dark 
matter and galaxy distributions and have found  that the null hypothesis is rejected only at 
the $23\%$ confidence level. 

\begin{figure}
  \begin{center}
   \plotone{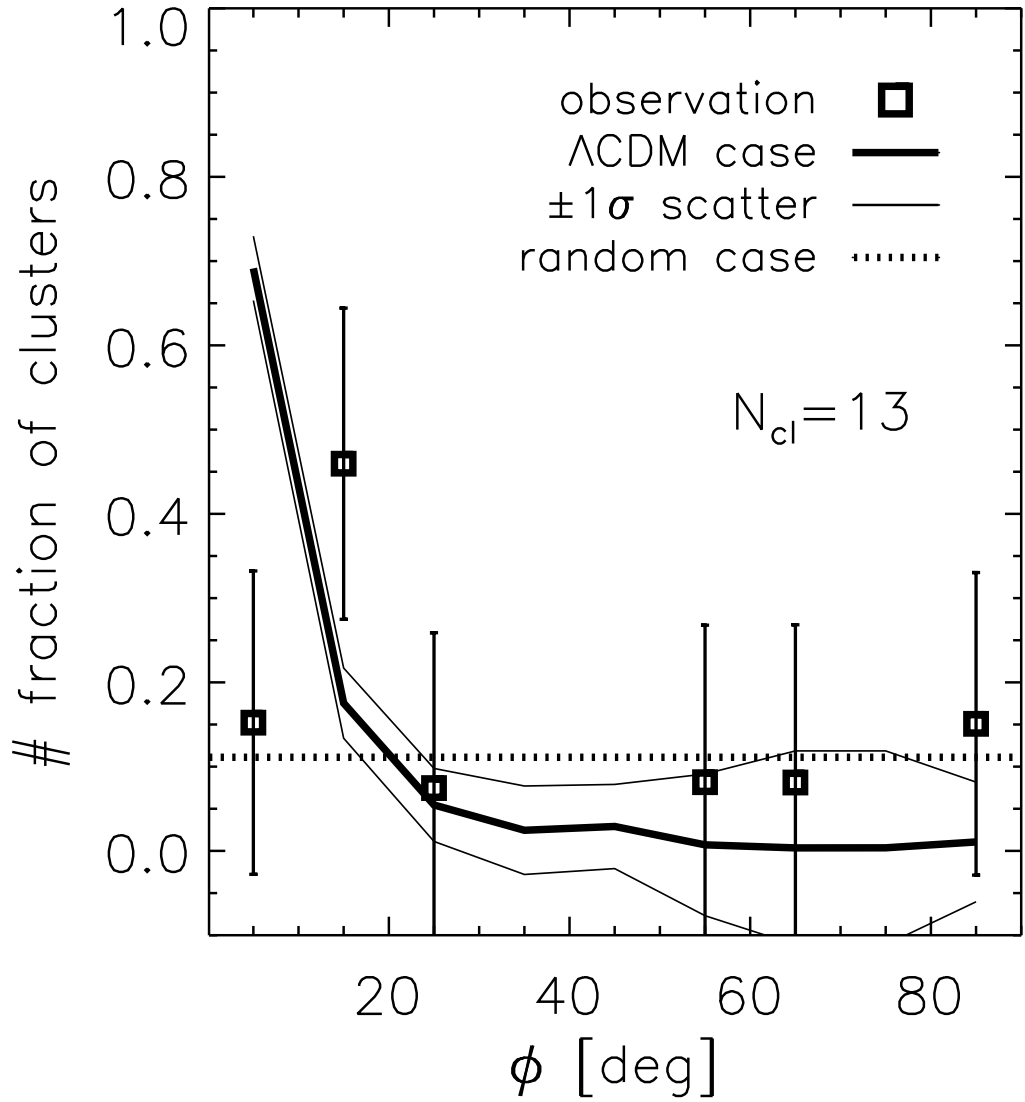}
\caption{Same as Figure \ref{fig:compare1} but including in the observational 
results only those clusters which do not show larger uncertainties in the 
measurement of the major axis directions.}
\label{fig:compare2}
 \end{center}
\end{figure}

Excluding those clusters whose position angles suffer from large uncertainties, 
we end up with a smaller subsample of $13$ clusters and redetermined the 
number fraction distribution as a function of $\phi$. Figure \ref{fig:compare2} 
plots the results. As it can be seen, the distribution has a higher peak 
at the second bin but still deviates significantly from the numerical result based on a 
$\Lambda$CDM cosmology and the semi-analytic galaxy formation model. 
The recalculation of $\chi^{2}$ for this case rejects the hypothesis that the 
observational result from the $13$ clusters is consistent with the numerical result  
is rejected at the $95\%$ confidence level.

It is worth mentioning the differences in the redshift range between the numerical and 
observational data used for the above comparison: The numerical result has been obtained 
at $z=0$ while the observational results have been drawn from the galaxy clusters at 
$0.1<z<0.3$. This difference in the redshift range, however, would even worsen the 
disagreement between the numerical and the observational result for the following reason. 
As mentioned in \S 2 and as shown by N-body simulations 
\citep[e.g.,][]{BS05,bailin-etal05,atlay-etal06,LK06}, the correlations between the major 
axes of dark matter and satellite galaxy distributions tend to be stronger at higher 
redshifts. In other words, the predicted strength of the correlations used for the 
comparison with the observational result is underestimated. 

One might think that the selection bias in the measurements of galaxy 
distributions should be responsible for the disagreement between theory and 
observation since in the analysis of \citet{Oguri-etal10} only those bright 
galaxies with $mag > 22$ at r-band are used unlike in the numerical analysis. 
Recall, however, that we used the luminosity-weighted galaxy distributions 
in the numerical analysis to determine the major axes, which should minimize the 
expected selection bias. 

Another difference lies in the values of the cosmological parameters used for 
the mili-Millennium Run simulations. Especially, the value of $\sigma_{8}$ chosen 
by the Millennium Run simulation has been known to be higher than the 
WMAP value \citep{wmap7}. The correlations between dark matter and satellite 
galaxy distributions in triaxial clusters, however, are unlike to be 
significantly affected by the initial cosmological conditions, since it 
represents a nonlinear observable rather than a linear one. 

\section{DISCUSSION}
 
Now that we have found the observed misalignments of the projected major 
axes of dark matter and galaxy distributions in the galaxy clusters to be  
inconsistent with the numerical prediction based on a $\Lambda$CDM cosmology and 
the semi-analytic galaxy formation model, we would like to discuss the 
possible physical origin of the misalignments. A usual suspect would be
some hydrodynamic gas physics involved in the galaxy formation that was not 
included in the Millennium Run simulation and semi-analytic galaxy catalog.  
If this is really the case, then our result indicates that the semi-analytic galaxy 
formation model has to be improved to account for the observed deviations of the 
orientations of cluster galaxy distributions relative to that of dark matter distributions.
 
An intriguing new possibility is that the orientations of member galaxy distributions 
have deviated from that of dark matter distributions due to the existence of the fifth 
force that is non-zero only in massive halos.   In some dark energy model where the scalar 
field is coupled both to the dark matter particles and the second relativistic particle 
species, a fifth force is short ranged, having non-zero values only on the scales 
smaller than the typical cluster sizes ($\sim 1$Mpc) 
\citep[e.g.,][]{FP04,nusser-etal05}. 

According to recent simulation results 
\citep[e.g.,][]{nusser-etal05,keselman-etal09,baldi-etal10}, in these 
models with a local fifth force the cluster galaxies (i.e., the galactic halos in dense 
environments) tend to form earlier and evolve more nonlinearly than in the standard 
$\Lambda$CDM model.  In addition, a local fifth force also tend to affect the mutual interactions between the satellite galaxies, which would result in an internal tidal field 
different from the gravitational external tidal field. 
Whereas, the dark matter particles which constantly accrete onto the host clusters at  
relatively later epochs than the member galaxies along the primary filaments from the surroundings would be less affected by the existence of a local fifth force and thus would 
keep the memory of the external gravitational tidal fields. Overall, a fifth force would play a 
role of diminishing the expected strong alignments between dark matter and galaxy 
distributions in clusters.

This new idea also suggests that it might be possible to constrain the range and 
strength of a local fifth force by measuring the misalignments between dark matter and 
galaxy distributions in clusters.  This goal calls for making a quantitative prediction of the 
coupled dark energy model on the correlations between dark matter and galaxy distribution 
using detailed N-body and hydrodynamic simulations.

\acknowledgments

The Millennium Run simulation used in this paper was carried out by the Virgo 
Supercomputing Consortium at the Computing Centre of the Max-Planck Society 
in Garching, which are available at http://www.mpa-garching.mpg.de/millennium.
We thank M.Oguri for useful comments. This work was supported by the National 
Research Foundation of Korea (NRF) grant funded by the Korea government (MEST, 
No.2010-0007819).

\clearpage
\clearpage
\end{document}